# The scientific payload on-board the HERMES-TP and HERMES-SP CubeSat missions

Y. Evangelista[*a,b], F. Fiore[c], F. Fuschino[d,e], R. Campana[d,e], F. Ceraudo[a], E. Demenev[f], A. Guzman[g], C. Labanti[d], G. La Rosa[h], M. Fiorini[i], M. Gandola[j], M. Grassi[k], F. Mele[j], D. Milánkovich[l], G. Morgante[d], P. Nogara[h], A. Pál[m], R. Piazzolla[a], S. Pliego Caballero[g], I. Rashevskaya[n], F. Russo[h], G. Sciarrone[a], G. Sottile[h], F. Ambrosino[a], N. Auricchio[d], M. Barbera[o,p], P. Bellutti[f], G. Bertuccio[j], G. Borghi[f], J. Cao[q], T. Chen[q], G. Dilillo[r], M. Feroci[a,b], F. Ficorella[f], U. Lo Cicero[o,p], P. Malcovati[k], A. Morbidini[a], G. Pauletta[r], A. Picciotto[f], A. Rachevski[s], A. Santangelo[g], C. Tenzer[g], A. Vacchi[r], L. Wang[q], Y. Xu[q], G. Zampa[s], N. Zampa[r,t], N. Zorzi[f], L. Burderi[u], M. Lavagna[v], R. Bertacin[w], P. Lunghi[v], A. Monge[x], B. Negri[w], S. Pirrotta[w], S. Puccetti[w], A. Sanna[u], F. Amarilli[y], G. Amelino-Camelia[z], M. Bechini[v], M. Citossi[r], A. Colagrossi[v], S. Curzel[v], G. Della Casa[r], M. Cinelli[ii], M. Del Santo[h], T. Di Salvo[o], A. Anitra[o], C. Feruglio[c], F. Ferrandi[aa], M. Fiorito[v], D. Gacnik[bb], G. Galgóczi[cc], A.F. Gambino[o], G. Ghirlanda[dd], A. Gomboc[ee], M. Karlica[ee], P. Efremov[ee], U. Kostic[ff], A. Clerici[ff], B. Lopez Fernandez[x], A. Manca[u], A. Maselli[w], L. Nava[c], M. Ohno[cc], D. Ottolina[v], A. Pasquale[v], M. Perri[gg,w], M. Piccinin[v], J. Prinetto[v], A. Riggio[u], J. Ripa[cc,hh], A. Papitto[gg], S. Piranomonte[gg], F. Scala[v], D. Selcan[bb], S. Silvestrini[v], T. Rotovnik[bb], I. Troisi[v], E. Virgilli[d], N. Werner[hh,cc] and G. Zanotti[v]

[a]INAF-IAPS Rome, Italy; [b]INFN sez. Roma Tor Vergata, Italy; [c]INAF-OATS, Trieste, Italy; [d]INAF-OAS Bologna, Italy; [e]INFN sez. Bologna, Italy; [f]Fondazione Bruno Kessler, Trento, Italy; [g]IAAT University of Tuebingen, Germany; [h]INAF-IASF Palermo, Italy; [i]INAF-IASF Milano, Italy; [j]Politecnico di Milano, Department of Electronics, Information and Bioengineering, Como, Italy; [k]University of Pavia, Department of Electrical, Computer, and Biomedical Engineering, Pavia, Italy; [l]C3S Electronics Development LLC, Budapest, Hungary; [m]Konkoly Observatory, Budapest, Hungary; [n]TIFPA-INFN, Trento, Italy; [o]Università degli Studi di Palermo, Dipartimento di Fisica e Chimica Emilio Segrè, Palermo, Italy; [p]INAF-OAP Palermo, Italy; [q]Institute of High Energy Physics, Chinese Academy of Sciences, Beijing, China; [r]Università di Udine; [s]INFN sez. Trieste, Italy; [t]INFN sez. Udine, Italy; [u]Dipartimento di Fisica Università degli Studi di Cagliari, Italy; [v]Dipartimento di Scienza e Tecnologia Aerospaziali, Politecnico di Milano, Italy; [w]Agenzia Spaziale Italiana, Italy; [x]DEIMOS, Spain; [y]Fondazione Politecnico Milano, Italy; [z]Università Federico II Napoli, Italy; [aa]Dipartimento di Informatica, Politecnico di Milano, Italy; [bb]Skylabs, Slovenia; [cc]ELTE, Hunghary; [dd]INAF-OAB Brera, Milano, Italy; [ee]University of Nova Gorica, Slovenia; [ff]Aalta Lab, Slovenia; [gg]INAF-OAR Roma, Italy; [hh]Department of Theoretical Physics and Astrophysics, Masaryk University, Brno, Czech Republic; [ii]Dipartimento di Matematica, Università di Roma Tor Vergata, Italy;

## ABSTRACT

HERMES (*High Energy Rapid Modular Ensemble of Satellites*) Technological and Scientific pathfinder is a space borne mission based on a LEO constellation of nano-satellites. The 3U CubeSat buses host new miniaturized detectors to probe the temporal emission of bright high-energy transients such as Gamma-Ray Bursts (GRBs). Fast transient localization, in a field of view of several steradians and with arcmin-level accuracy, is gained by comparing time delays among the same event detection epochs occurred on at least 3 nano-satellites. With a launch date in 2022, HERMES transient monitoring represents a keystone capability to complement the next generation of gravitational wave experiments. In this paper we will illustrate the HERMES payload design, highlighting the technical solutions adopted to allow a wide-energy-band and sensitive X-ray and gamma-ray detector to be accommodated in a CubeSat 1U volume together with its complete control electronics and data handling system.

---

[*]yuri.evangelista@inaf.it; hermes-sp.eu

## 1. INTRODUCTION

HERMES (*High Energy Rapid Modular Ensemble of Satellites*) is a mission concept based on a constellation of nano-satellites (3U CubeSats) in low Earth orbit, hosting innovative X and γ-ray detectors to probe the high-energy emission of bright transients [1][2][3]. HERMES is built upon a twin project: the HERMES Technological Pathfinder (HERMES-TP), funded by the Italian Ministry for education, university and research and the Italian Space Agency, and the HERMES Scientific Pathfinder (HERMES-SP), funded by the European Union’s Horizon 2020 Research and Innovation Programme under Grant Agreement No. 821896.

Both projects (HERMES-TP and HERMES-SP) provide three complete satellites (payload and service module) to the constellation, aiming at demonstrating that fast GRB detection and localization is feasible with disruptive technologies on-board miniaturized spacecrafts, mostly exploiting commercial off-the shelf (COTS) components at a cost 1–2 order of magnitude lower than that of standard space projects (e.g. ESA M-class missions and NASA Explorer missions) and with a development time of a just a few years. Moreover, the Italian Space Agency approved and funded the participation to the SpIRIT (*Space Industry – Responsive – Intelligent – Thermal Nanosatellite*) CubeSat. The SpIRIT project, which is supported by the Australian Space Agency and led by University of Melbourne, will host an HERMES-like detector thus providing a seventh unit to the HERMES constellation.

The HERMES scientific goal is to demonstrates the accurate and prompt localization of bright hard X-ray/soft gamma-ray transients such as Gamma-Ray Bursts (GRBs). Fast high-energy transients are among the likely electromagnetic counterparts of the gravitational wave events (GWE) recently discovered by Advanced LIGO/Virgo [4], and of the Fast Radio Bursts. Moreover, HERMES will open the window of X-ray timing down to a fraction of microseconds, and thus investigate for the first time the temporal structure of GRBs at this scale to constrain models for the GRB engine and to test quantum space-time scenarios by measuring the delay time between GRB photons of different energy.

Being based on cost-effective nanosatellites, HERMES is intrinsically a modular project. This allows to avoid single or even multiple point failures (if one or several units are lost the constellation and the experiment is not lost) and to in-flight test the hardware and on-board software with the first launches. If needed, both hardware and software can be improved with the following launches.

## 2. HERMES PAYLOAD REQUIREMENTS

The flow-down of HERMES-TP and HERMES-SP Scientific Requirements [1] results in ambitious payload (P/L) requirements: broad energy band, good efficiency, good energy resolution, high temporal resolution, extremely compact design, reliable operation in a quite broad range of space environments (e.g., temperature, radiation damage, etc.). Although these goals are certainly reachable on standard scientific experiments, an innovative and customized payload design has been implemented in HERMES-TP/SP in order to fulfil the requirements in the limited mass, volume and power resources of a CubeSat.

Given the required broad energy range, an integrated detector [5] has been designed to exploit a “double detection” mechanism, with a partial overlap of the two detection systems around ~20 keV. Detection of soft X-rays (hereafter X-mode) is obtained by a segmented solid state detector employing custom designed Silicon Drift Detector (SDD) [6], with a cell size of about 0.45 cm$^2$, which allows to attain a low noise level (of the order of a few tens of e$^-$ rms at room temperature) and correspondingly a low energy threshold for the detection of X-ray radiation (<5 keV). On the other hand, detection of hard X-rays and γ-rays (hereafter S-mode) is obtained with a two-stage process, which converts high-energy photons into visible light by means of scintillator crystals; optical photons are then collected and converted into electric charge in the same photodetector (SDD) used to directly detect X-ray photons. In this case, the SDDs act as a photodiode and produces an amplitude charge signal proportional to the amount of scintillator light collected.

The discrimination between the two signals in the SDD (soft X-rays or optical photons) is achieved using the segmented design: each scintillator crystal is read-out by two SDD cells, so that events detected in only one SDD are associated to soft x-rays converted in a single SDD cell, while events detected simultaneously in more than one SDD are due to optical light produced in the scintillator by an incoming hard X-ray/γ-ray. Table 1 summarizes the main HERMES-TP/SP Payload Requirements.

Table 1 HERMES-TP/SP Payload requirements

| Requirement | Condition | Value |
|---|---|---|
| **Sensitivity** | $E \leq 20$ keV (GRB short/long) | $\leq 2$ photons/s/cm$^2$ |
| | $50 \leq E \leq 300$ keV (GRB short/long) | $\leq 1$ photons/s/cm$^2$ |
| **Energy band** | $E_{low} \leq 5$ keV, $E_{high} \geq 500$ keV | |
| **Peak effective area** | X-mode | $\geq 50$ cm$^2$ |
| | S-mode | $\geq 50$ cm$^2$ |
| **Lower energy threshold** | | $\leq 5$ keV |
| **Energy resolution EOL** | 5.0 – 6.0 keV | $\leq 1$ keV FWHM |
| | 50.0 – 60.0 keV | $\leq 5$ keV FWHM |
| **Time resolution (1σ)** | X-mode | $\leq 400$ ns |
| | S-mode | $\leq 250$ ns |
| **Time accuracy (1σ)** | GPS locked | $\leq 100$ ns |
| | GPS unlocked (up to 1500 s) | $\leq 200$ ns |
| **Field of view** | | $\geq 3$ sr FWHM |
| **Background rate 50–300 keV** | | $\leq 1.5$ counts/s/cm$^2$ |
| **Background rate 20–300 keV** | | $\leq 12$ counts/s/cm$^2$ |
| **Background knowledge** | | $\leq 5\%$ |
| **Maximum sustainable GRB flux** | | 40000 counts/s |
| **On-board memory** | | $\geq 16$ Gbit |
| **Mass allocation** | | $< 1.8$ kg |
| **Volume allocation** | | $\leq 10 \times 10 \times 12.5$ cm$^3$ (1.25 U) |
| **Power allocation** | | $\leq 5$W |
| **Detector operative temperature range** | | –30 °C ÷ +10 °C |
| **P/L non operative temperature range** | | –40 °C ÷ +80 °C |

## 3. HERMES PAYLOAD DESIGN

Figure 1 shows an exploded view of the HERMES-TP/SP payload. The payload is composed by 4 main subsystems: the detector assembly (DA), the back-end electronic board (BEE), the power-supply electronic board (PSU) and the payload data handling unit (PDHU). In the following we provide detailed description of each payload subsystem.

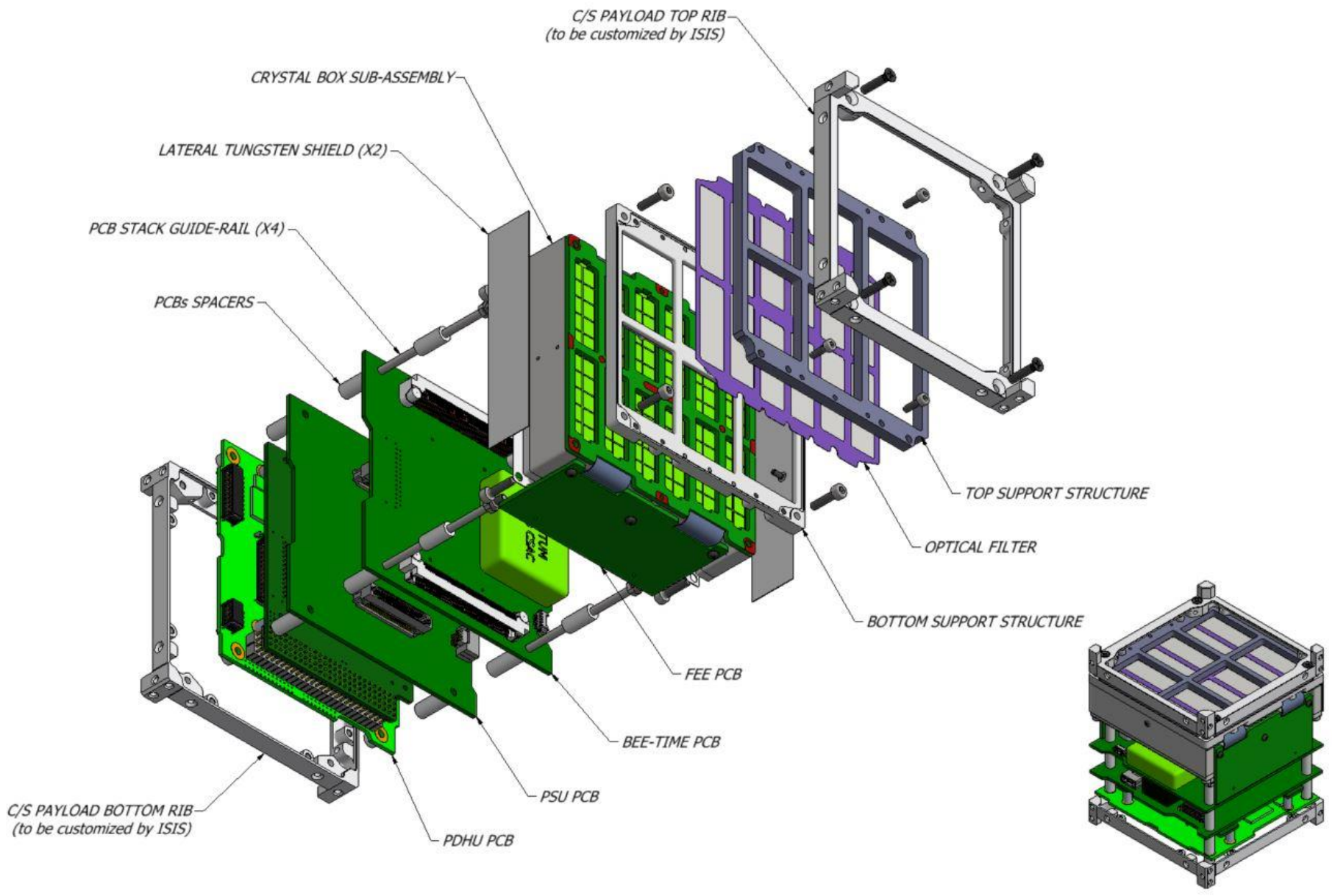

Figure 1 Exploded view of the HERMES-TP/SP P/L

## 3.1 Detector Assembly (DA)

The HERMES-TP/SP detector assembly [5] is composed of:

- a **detector support structure**, mounted on the top of the sensitive plane. This structure is split in an upper and a lower part, with an optical filter in-between;
- a **front-end electronics** (**FEE)** board, a rigid-flex printed circuit board with a slotted main plate and two side wings, connected to the main part with a flexible flat cable integrated in the board.
  The FEE PCB hosts:
  - 12 Silicon Drift Detector (SDD) arrays, each one with 10 independent cells, for a total of 120 detector channels;
  - 120 LYRA-FE ASIC dies, one for each SDD channel;
  - 4 LYRA-BE ASICs, each one interfaced with 30 LYRA-FE channels;
  - discrete passive and active electronic components;
  - two electrical connectors (one per side) providing the electrical interfaces with the BEE board;
- 60 **scintillator crystals**, each one optically connected to two SDD cells. To maximize the optical contact between SDD and crystal, ensuring optimal scintillation light readout, each crystal is:
  - o optically isolated on each side except the one in direct contact with the SDD, with a reflecting material;
  - o optically connected to the SDD through a transparent silicone layer;
  - o mechanically fixed inside satellite frame between the backside of the SDD box and the detector support structure;
- a **crystal box,** hosting the 60 scintillator crystal and providing the overall DA mechanical structure together with the FEE board and the detector support structure. Three 200 µm thick tungsten sheets are glued on the bottom side and on two lateral sides of the crystal box to provide background shielding.

Figure 2 shows an exploded view of the HERMES-TP/SP DA.

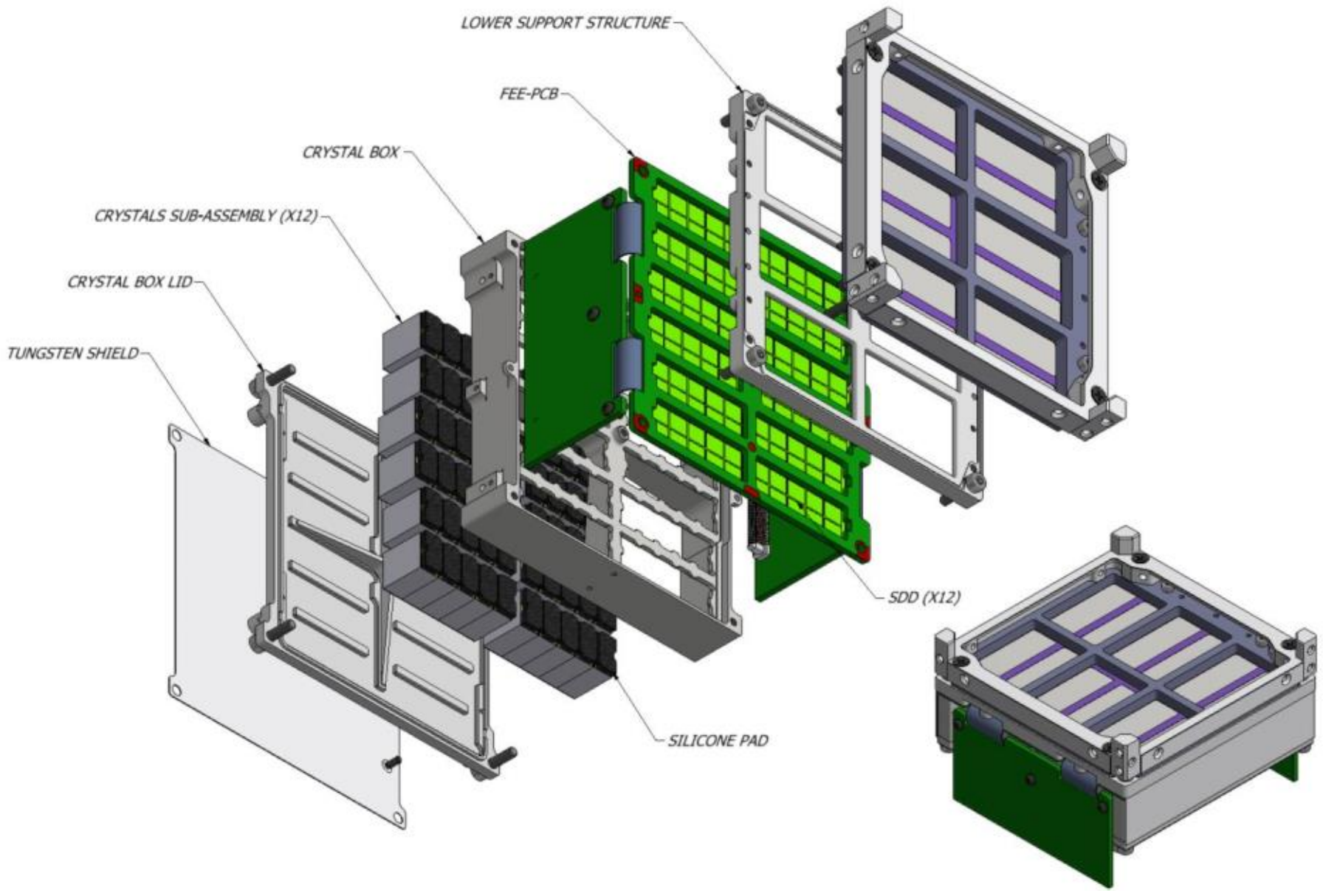

Figure 2 Exploded view of the HERMES-TP/SP Detector Assembly (DA)

As reported before, in order to reach a very wide energy band in an extremely compact instrument, the detector is based on the so-called “siswich” concept [7][8]. In this architecture, the SDDs play the double role of read-out device for the optical signal from the scintillator and of an independent X-ray solid state detector. The intrinsic low-noise characteristics of the silicon drift technology allows reaching a low (< 20–30 keV) energy scintillator threshold. Above these energies, the increasing sensitivity of the scintillator is able to compensate for the lack of efficiency of thin silicon sensors, so a reasonably flat efficiency in a wide energy band for the whole integrated system is reached.

The HERMES SDDs are based on the state-of-the-art results achieved within the framework of the Italian ReDSoX collaboration[1], with the combined design and manufacturing expertise of INFN-Trieste and Fondazione Bruno Kessler (FBK, Trento). Each SDD array (Figure 3) is composed by 2×5 cells on a 450 μm thick silicon substrate, with a cell of 7.44×6.05 $mm^2$ for a total of 4.5 $cm^2$ sensitive area. The whole matrix is surrounded by a 1.2 mm wide guard region which allows for a correct electric field termination. The overall geometric area of the SDD array is 39.6×14.5 $mm^2$. By exploiting the single-side biasing technique [9], all the SDD bonding pads are placed on the detector *n*-side (i.e., the anode side), thus providing a homogenous light entrance window for scintillation light readout. Four 500 μm wide metallization strips are present on the *p*-side (optical window side) to ensure negligible optical crosstalk of the scintillation light emitted by adjacent crystals.

[1] http://redsox.iasfbo.inaf.it

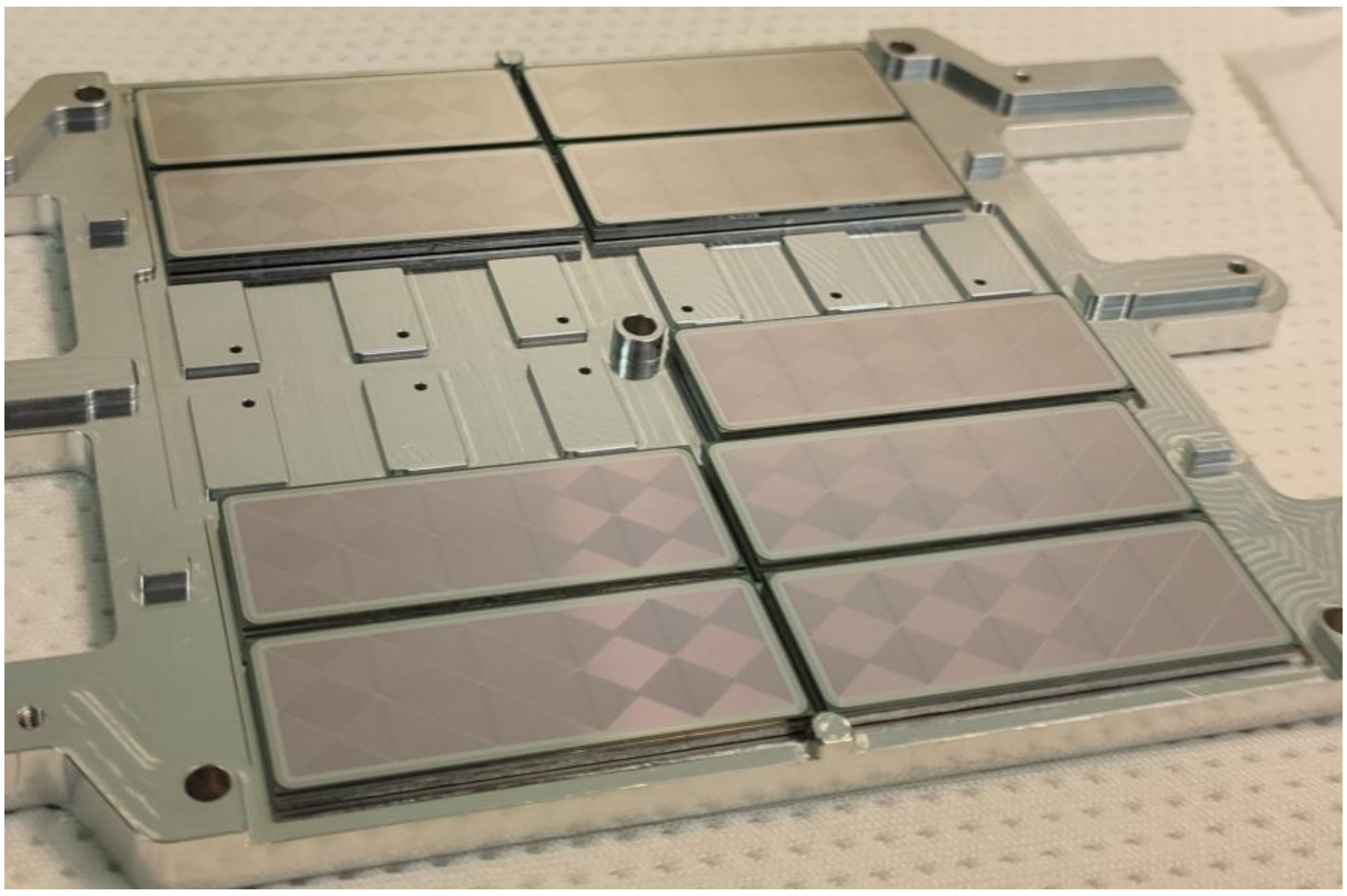

Figure 3 HERMES Silicon Drift Detectors (SDDs) on the detector integration jig

The scintillator are cerium-doped gadolinium-aluminum-gallium garnet crystals ($Gd_3Al_2Ga_3O_{12}$ or Ce:GAGG), developed firstly in Japan around 2010, and commercially available since 2014 [10][11]. Ce:GAGG scintillators have high intrinsic light output (~50000 photons/MeV), no intrinsic background, no hygroscopicity, fast radiation decay time of ~90 ns, high density (6.63 g/cm$^3$), peak light emission at 520 nm and high effective mean atomic number (54.4). All these characteristics make the Ce:GAGG scintillators the optimal choice for the HERMES detector. The crystals have a $6.94 \times 12.10$ mm$^2$ cross section and a 15 mm thickness, with chemically polished faces. Each crystal is optically coupled to two adjacent SDD cells by means of a 3.3 mm thick Polydimethylsiloxane encapsulant (DOWSIL 93-500), allowing for a reliable discrimination of X-ray and γ-ray photons via multiplicity analysis. The crystals are individually wrapped with a 3M DF2000MA specular reflector film and grouped in sub-assemblies of five crystals, with one sub-assembly coupled to an individual SDD array (Figure 4).

With the reported optical configuration, the crystal measured light output at room temperature is ~28 e$^-$/keV, which translates in ~14 e$^-$/keV collected by each SDD channel reading-out the scintillator.

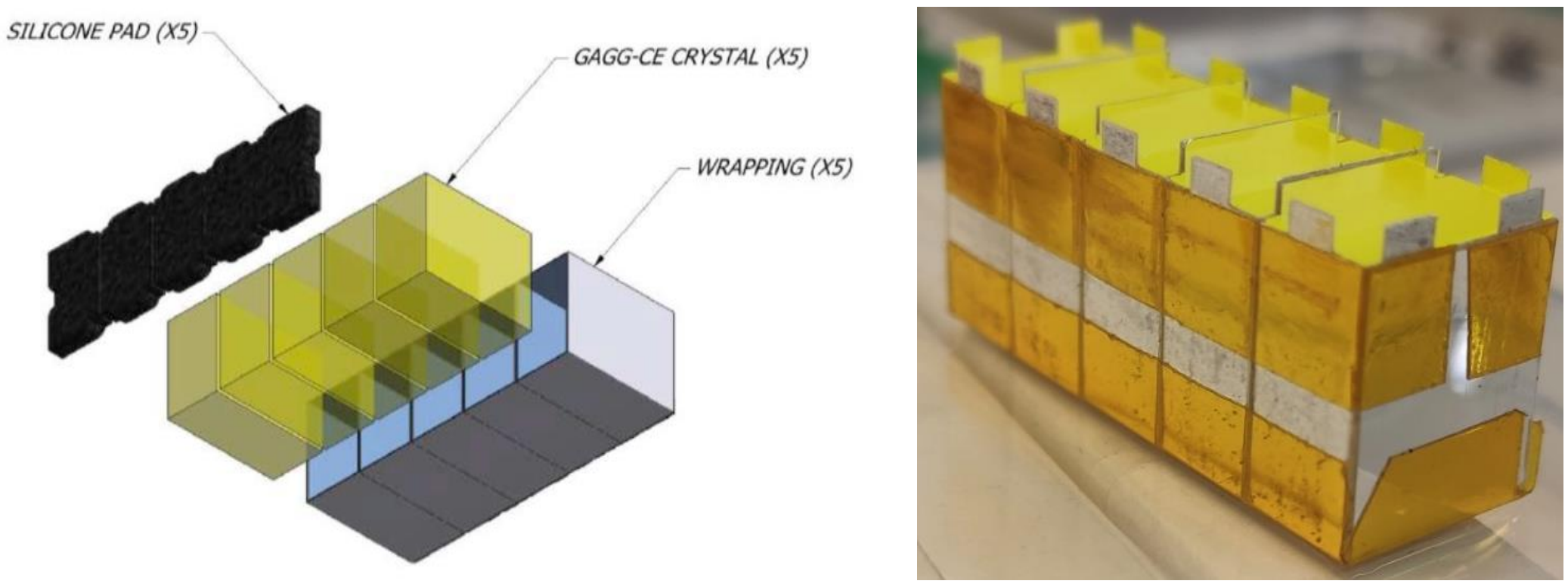

Figure 4 Ce:GAGG crystal optical assembly

The Front-end electronics board (FEE, Figure 5) accommodates the 12 SDD matrices, 120 LYRA-FE ASICs, 4 LYRA-BE ASICs, and several passive electronic components. The PCB is manufactured with a rigid-flex technology, allowing the bending of the two side wings. Each wing hosts 2 LYRA-BE ASICs and a board-to-board connector providing the FEE electrical interface with the back-end electronic board.

The 120 LYRA-FE chips are mounted on the top PCB layer, while the 12 SDD arrays are glued on the bottom PCB layer by means of DOWSIL 3145 silicone adhesive. Aluminum wire bondings provide the electrical connection between the ASICs and the FEE PCB (chip-to-board), the SDD biases (chip-to-board), and connection between the SDD anodes and the LYRA-FE input pads (chip-to-chip).

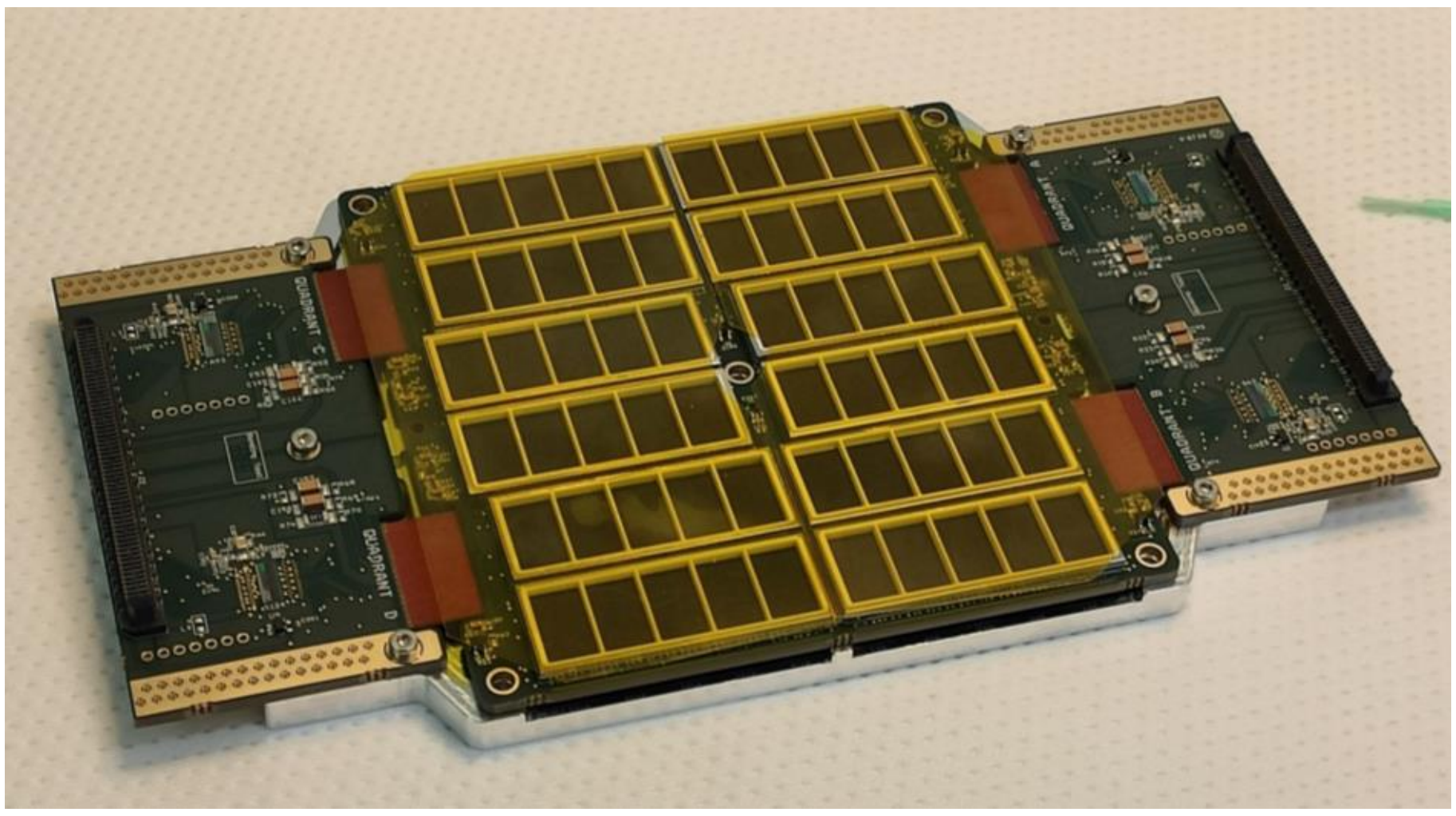

Figure 5 Demonstration model of the HERMES-TP/SP front-end PCB (FEE), integrated with 12 SDD arrays and LYRA-FE and LYRA-BE ASICs. See text for more details

The read-out ASIC design is based on the heritage of the VEGA ASIC [12][13], developed by Politecnico of Milano and University of Pavia for the readout of large area SDD in the framework of the LOFT Phase-A study (ESA M3 Cosmic Vision program).

In order to optimize the system noise performance, a distributed architecture has been adopted for the HERMES ASICs, specifically designed and manufactured in AMS 0.35-µm CMOS mixed signal technology (Figure 6). The Front-End ICs (LYRA-FE), which include the preamplifier, the first shaping stage and signal line-transmitter, are placed as near as possible to each SDD anode. The current signal produced by the first-stage pulse shaper is transferred to one-channel of the back-end IC (LYRA-BE), where it is collected by the current receiver block and further processed. Each LYRA-BE can manage the signals produced by 30 LYRA-FE ICs, corresponding to one fourth of the HERMES-TP/SP detection plane. Detailed description of the LYRA-FE and LYRA-BE ASICs can be found in [14][15].

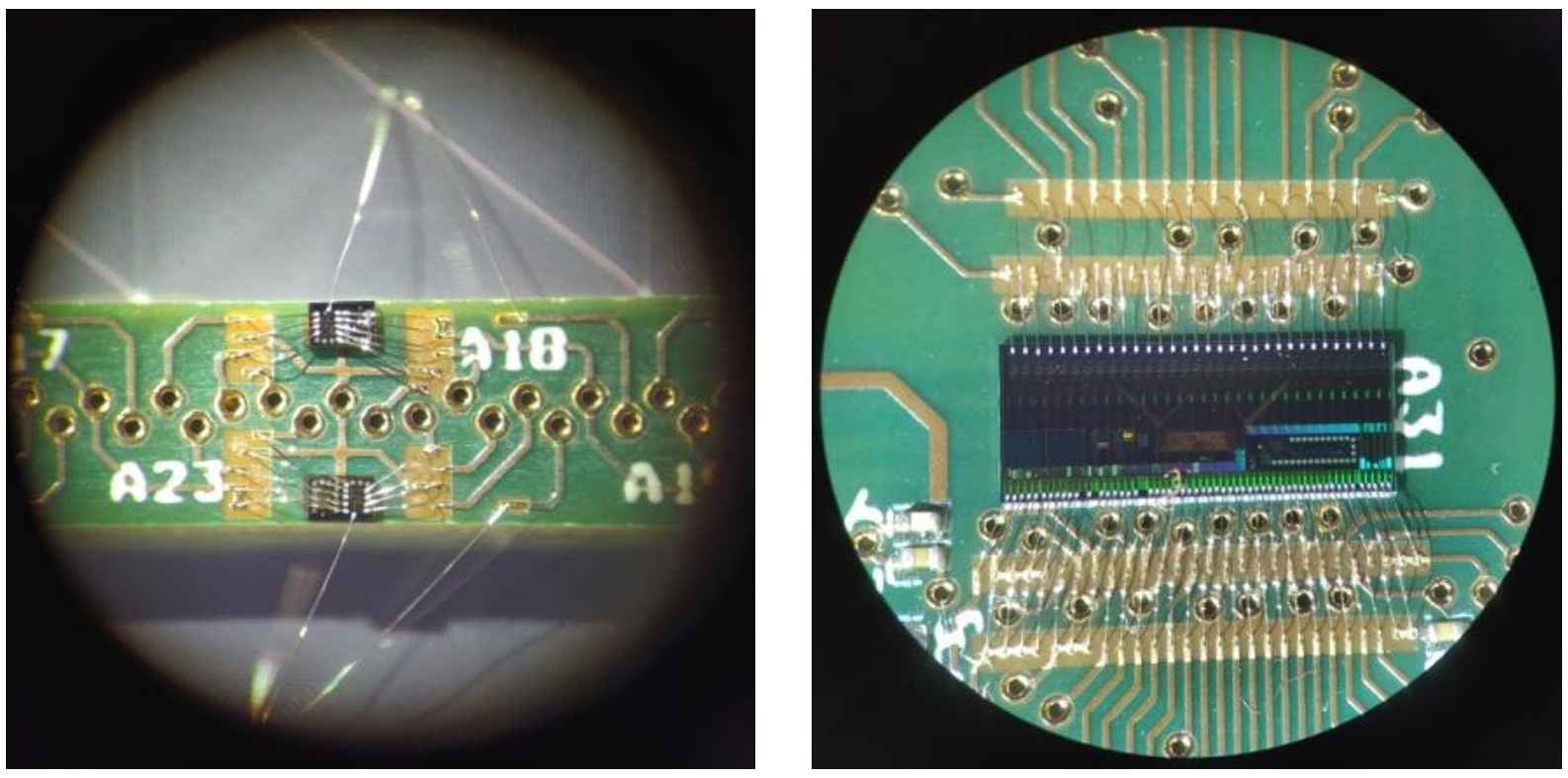

Figure 6 LYRA-FE ASICs (left) and LYRA-BE ASIC (right) integrated on the HERMES-TP/SP FEE PCB

The HERMES-TP/SP detector assembly is completed by an optical filter and an AISI 316 stainless steel structure (Figure 7), which ensures the required mechanical stiffness to the FEE, encloses the Ce:GAGG crystals, and provide background shielding (photons and particles) by means of 200 µm thick tungsten sheets glued on the bottom and on two sides of the box containing the crystals. The top surface of each crystal is precisely aligned and held in position by means of tips featured at the upper surface of the crystal box. Moreover, in order to compensate the CTE mismatch between scintillators and crystal box and to withstand the dynamic forces acting on the scintillators during the ascending phases, each crystal is preloaded with a 0.5 mm thick silicone layer placed between the scintillator bottom face and the crystal box bottom lid.

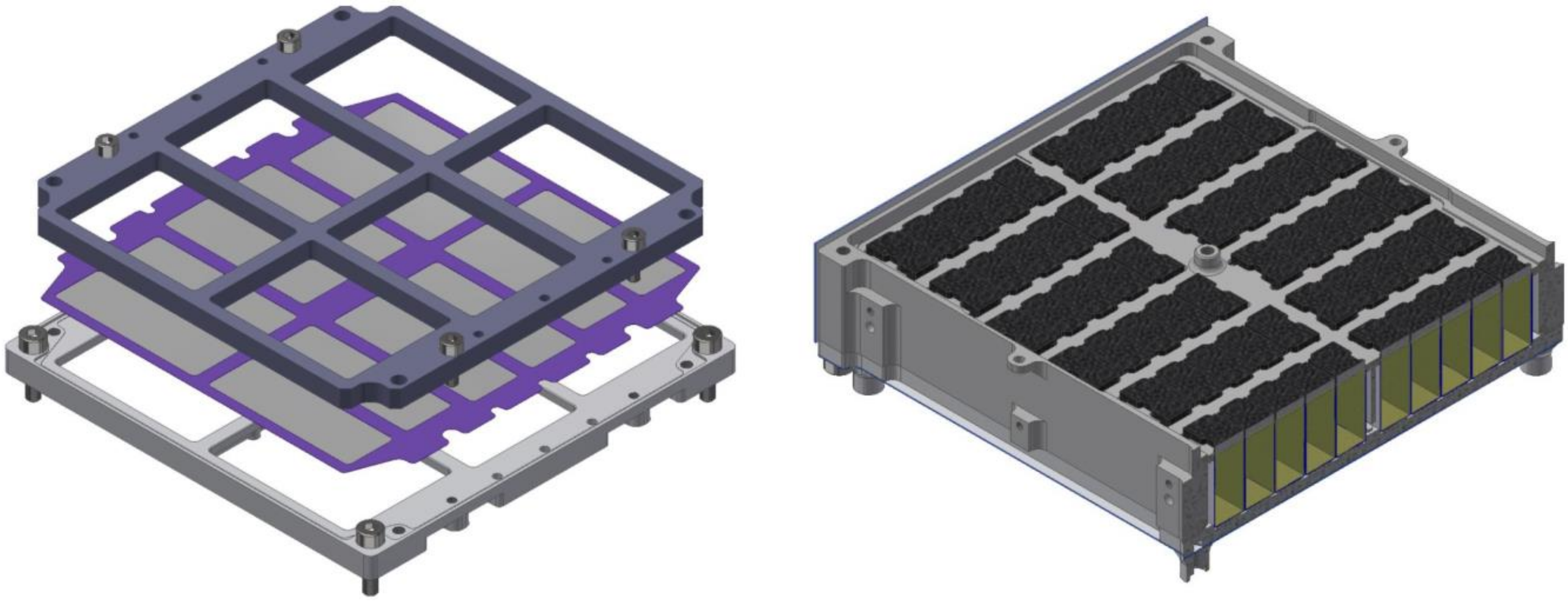

Figure 7 Left: Detector support structures (grey) and optical filter (purple). Right: Crystal box sub-assembly with wrapped scintillators and DOWSIL 93-500 optical (top) and preload pads (bottom)

The optical filter of the HERMES-TP/SP detector is part of the optical design of the P/L. Its prime task is to prevent NIR/O/UV light from reaching the SDD detector so to minimize the current noise generated by background photons with energies larger than the Silicon bandgap. In addition, being the P/L thermal design based on passive cooling only, the filter also enters the overall thermal design of the detector assembly. The HERMES-TP/SP filter has been manufactured by the Institute of High Energy Physics (IHEP) of the Chinese Academy of Sciences (CAS) and consists of a 1 µm polyimide layer with 200 nm Aluminum filming. A 100 µm thick nickel frame provide the filter mechanical stiffness and the mounting interface with the detector support structure. As shown in Figure 8, a thickness of 1 µm for the Polyimide and an Aluminum layer of 200 nm are acceptable from the point of view of transparency to X-rays, while guaranteeing the required $10^{-7}$ absorption in the $10^{2}$ nm – $10^{5}$ nm wavelength range.

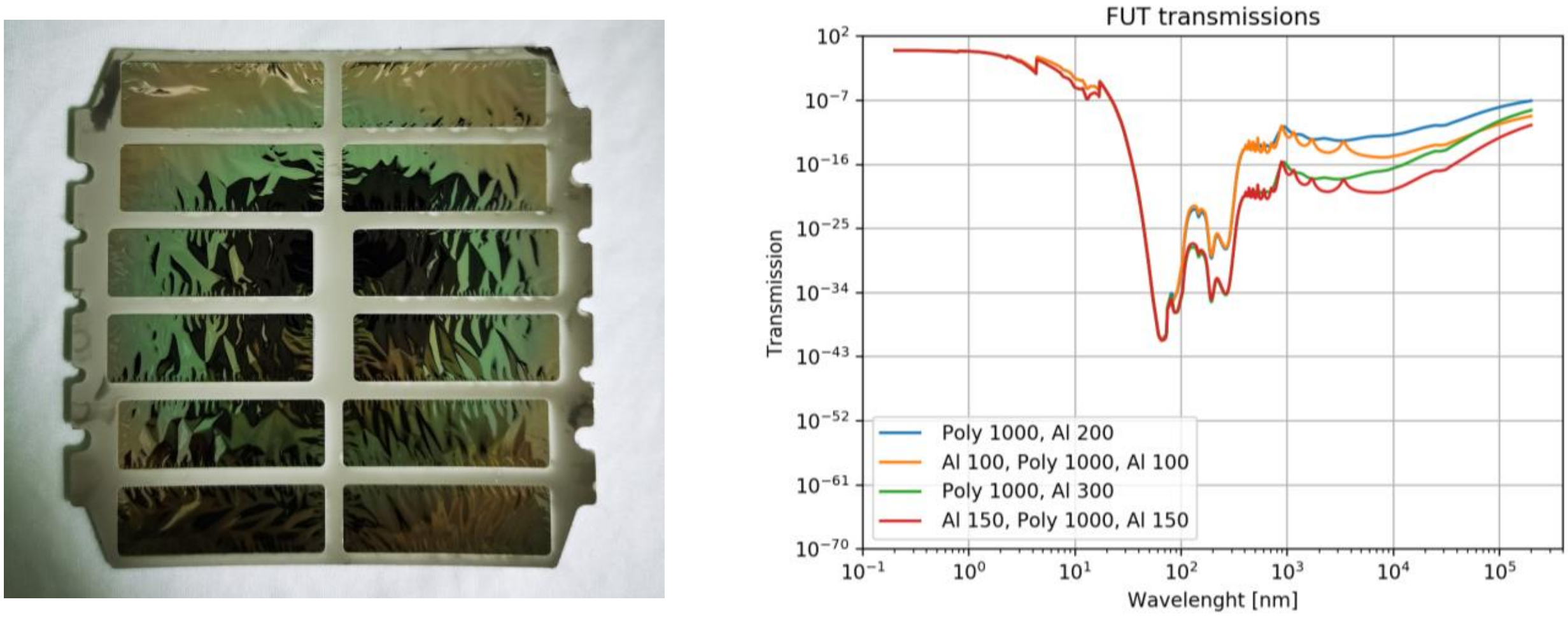

Figure 8 Left: Demonstration model of the HERMES-TP/SP filter manufactured by IHEP (Beijing). Right: Soft-X/UV/O/IR transmission of the optical filter estimated with different thickness of polyimide (1 µm and 2 µm) and different total thickness of Al coating (200 nm and 300 nm) deposited on one or two sides of the polyimide substrate

### 3.2 Back-end electronic board (BEE)

The BEE (Figure 9) is the logic block between the front-end ASICs and the Payload Data Handling Unit. The BEE is in charge of managing the ASICs configuration, the analog to digital conversion of the ASIC signals and the event time-tagging exploiting the sub-microsecond accuracy of a local chip-scale atomic clock. Moreover, the BEE collects the detector house-keeping data, commands the power lines required by the FEE, manages events data acquisition, transmits science data and HKs to the PDHU as well as to receive and decode tele-commands sent by the PDHU.

The core of the BEE is a SEL immune Intel/Altera Cyclone V FPGA (5CEFA4F23I7N) that implements all the required functions and tasks.

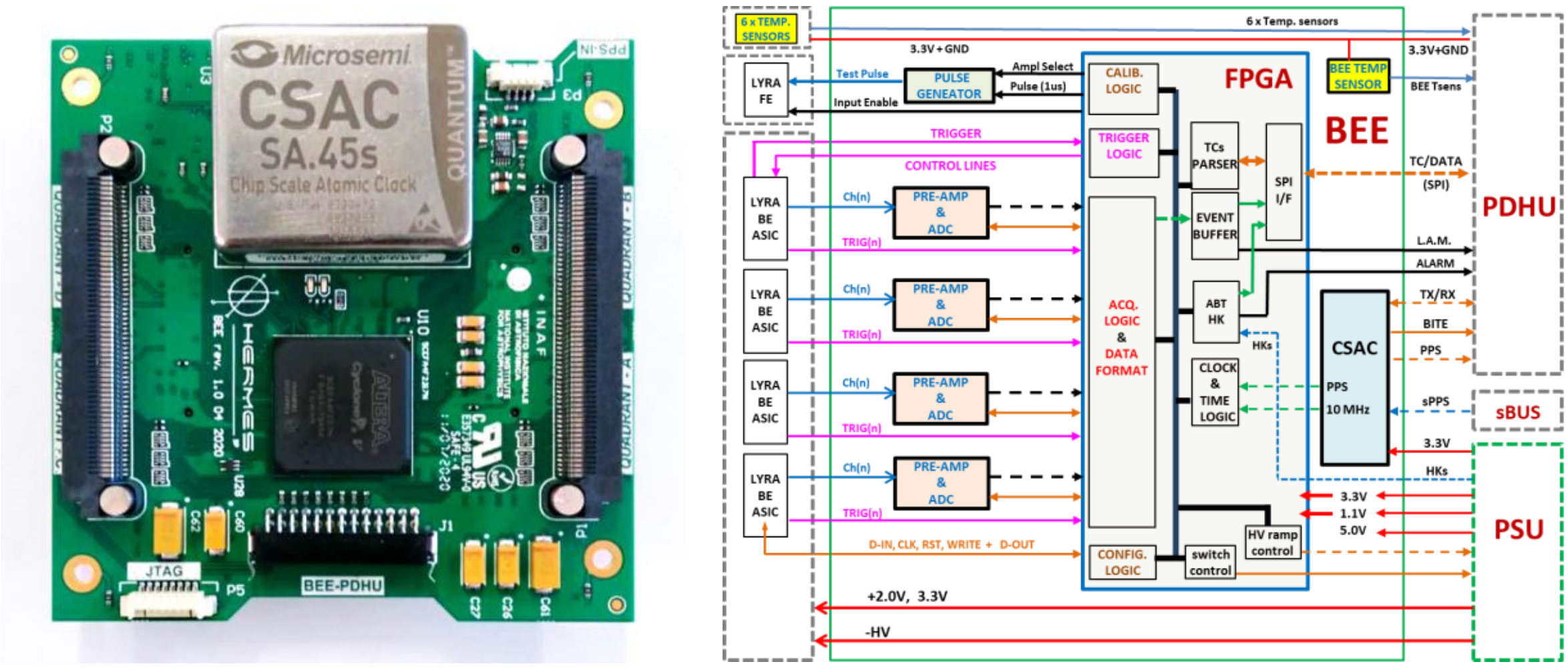

Figure 9 Left: Demonstration model of the HERMES-TP/SP back-end electronics PCB. Right: schematic block diagram of the BEE

The BEE receives, from the LYRA-BE ASICs, the analog event signals produced by the detectors and processed by the front-end board electronics. The signal coming from each LYRA-BE is routed through a preamplifier/buffer to an independent 12 bit, 1 MHz ADC, thus allowing for the simultaneous acquisition of the events detected in different quadrants. With this architecture, the four detector quadrants behave as independent instruments, thus providing also redundancy to the instrument.

The BEE communicates with the PDHU through a serial line (SPI). All data exchanges are initiated by the PDHU (master) through telecommands for writing, reading and task execution. The BEE receives and verifies the command and executes the function commanded. The configuration of the LYRA-BE ASICs and of the payload operational modes are transmitted by the PDHU under specific write commands. All the data acquired by the BEE (both scientific events and housekeeping) are saved in a temporary buffer and transmitted to PDHU once per second. In case of substantial buffer occupation, for example in presence of a high intensity astrophysical event, the BEE request an immediate data transmission to the PDHU by activating a dedicated look-at-me digital line.

To generate the time reference for events time-tagging a combination of GPS Pulse per Second signal (PPS) from the HERMS service module and a local ultra-stable clock is used. The ultra-stable clock is the space-qualified Microsemi SA.45s chip-scale atomic clock (CSAC), which provides 1PPS and 10 MHz clock output signals with an accuracy of ±5.0E-11, a short-term Allan deviation lower than $10^{-11}$ (1000 s) and a typical aging rate of $9\times10^{-10}$/month. The BEE FPGA implements 2 counters: one incremented by the CSAC PPS (28 bits) and the other incremented by the 10 MHz clock (24 bits). When the GPS signal is locked, the CSAC PPS signal is synchronized every second to the rising edge of the PPS provided by the satellite bus, thus ensuring a continuous correction for the payload on-board time. In case of GPS not locked, for example when the number of the GPS satellites tracked by the satellite bus GPS receiver is lower than 4 due to the earth occultation, the CSAC still provides the 1PPS and 10 MHz clock signal in a free-running mode.

Exploiting this architecture, the HERMES-TP/SP expected timing performance are summarized in Table 2 and Table 3 for the GPS locked and unlocked cases respectively. For the GPS unlocked case (Table 3) an interval of 1500 s from the last GPS lock has been considered.

Table 2 Time precision budget for the GPS locked case

| **Mode** | **Time accuracy (68% c. l.)** | **Time resolution (68% c. l.)** | **Total (68% c. l.)** |
|---|---|---|---|
| X-Mode | 53.4 ns | 320 ns | 324 ns |
| S-mode | 53.4 ns | 216 ns | 222 ns |

Table 3 Time precision budget for the GPS unlocked case

| **Mode** | **Time accuracy (68% c. l.)** | **Time resolution (68% c. l.)** | **Total (68% c. l.)** |
|---|---|---|---|
| X-Mode | 181 ns | 320 ns | 368 ns |
| S-mode | 181 ns | 216 ns | 282 ns |

### 3.3 Power Supply Unit (PSU)

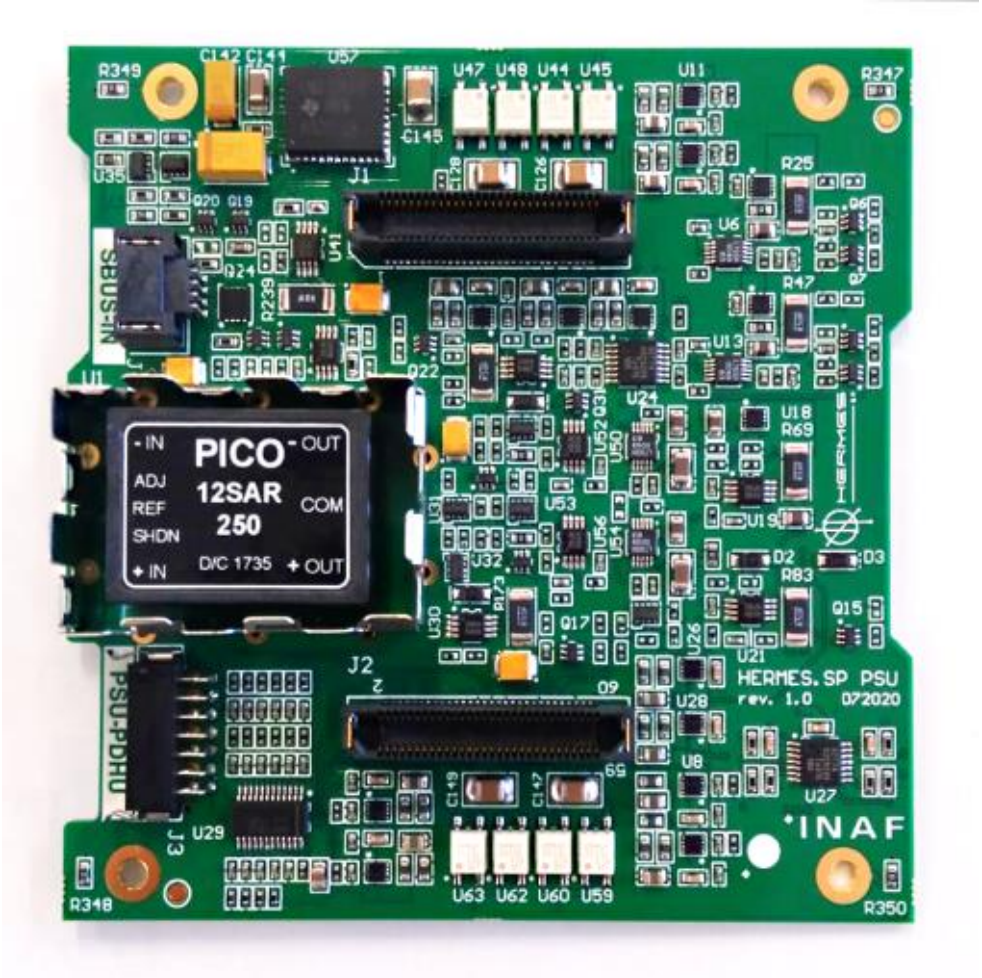

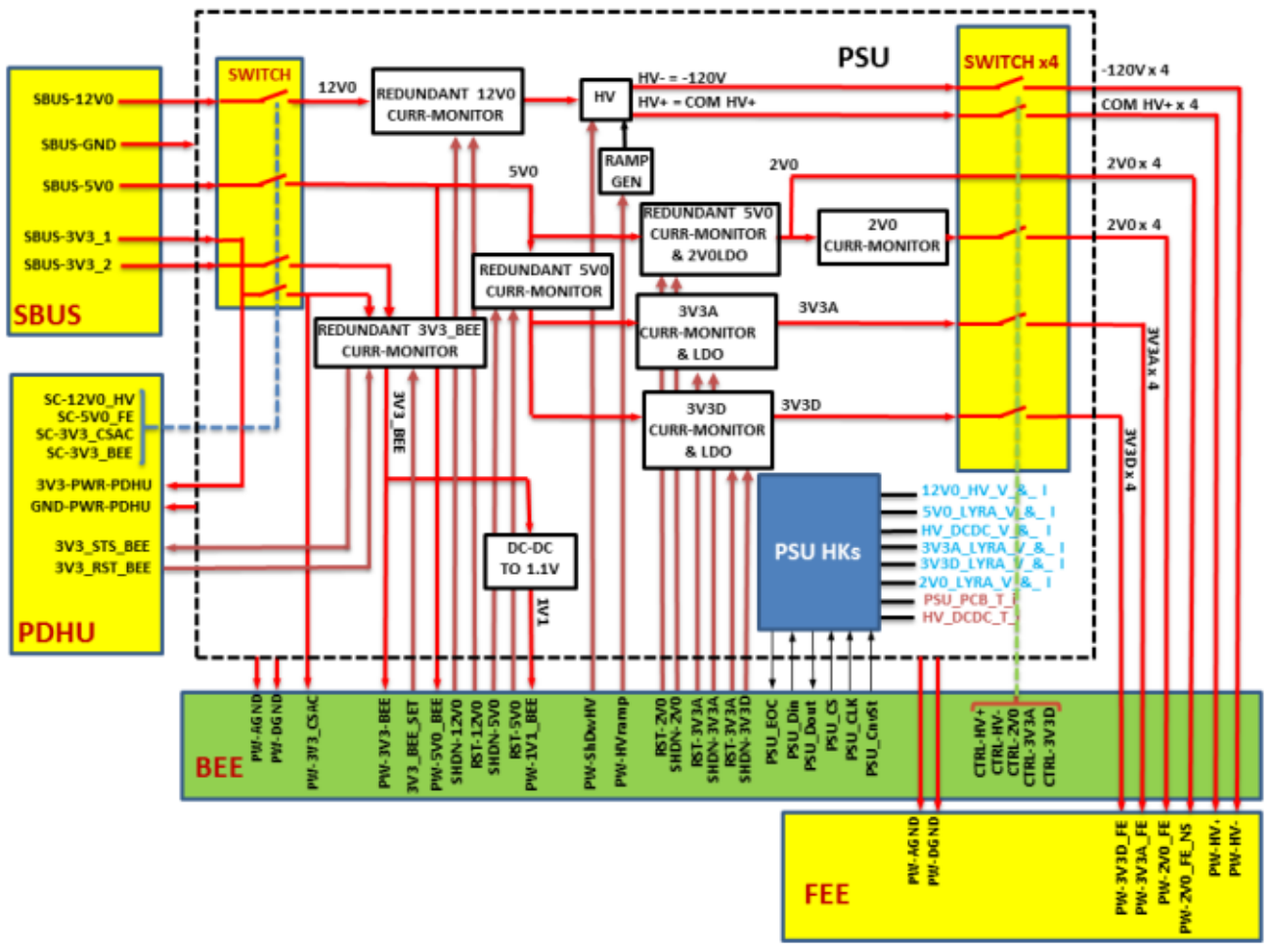

Figure 10 Left: Demonstration model of the HERMES-TP/SP Power Supply Unit (PSU). Right: schematic block diagram of the PSU

A custom Power Supply Unit (PSU, Figure 10) board has been designed and manufactured to provide the power supplies required by the payload (Figure 10). All the low voltages needed for the operation of the FEE are generated by ultra-low noise linear Low Drop-Out (LDO) regulators. A DC-DC converter (Picoelectronics 12SAR250) is used for the generation of the high-voltage detector biases while a Texas Instrument LMZ30602 is in charge of the generation of the 1.1V required by the BEE FPGA core.

The PSU receives from the satellite bus four power lines (12 V/80 mA, 5 V/100 mA and 2×3.3 V/1 A). Each line is routed through electronic switches placed on the PSU board and commanded by the PDHU. Another set of switches are placed in series with the main one and commanded by the BEE FPGA firmware.

All the low voltages lines are protected against overcurrent or latch-up in the powered circuitry by means of current monitors which also provide alert lines routed to the BEE FPGA for latch-up monitoring and logging. Special care has been taken in the PSU design to ensure a robust and safe latch-up control. In particular, a delay circuit is used to avoid the protection activation due to inrush current, and two of sense amplifiers are implemented in parallel in a redundant configuration for high reliability protection of sensitive power lines. Moreover, all the sense amplifiers are powered downstream the sense resistor.

### 3.4 Payload Data Handling Unit (PDHU)

The HERMES-TP/SP Payload Data Handling Unit (PDHU) is the interface between the spacecraft and the payload. The selected hardware for the PDHU is the Innovative Solutions In Space (ISIS) On-Board computer (iOBC). The iOBC is a flight proven, high performance processing unit based around an ARM9 processor, and offers a multitude of standardized interfaces. Combined with its daughterboard architecture, it allows for easy addition of mission specific electronics or interfaces. A custom-made daughterboard is the current baseline for the PDHU system and provides all the payload-bus electrical interfaces as well as the PHDU internal interface with the other payload subsystems.

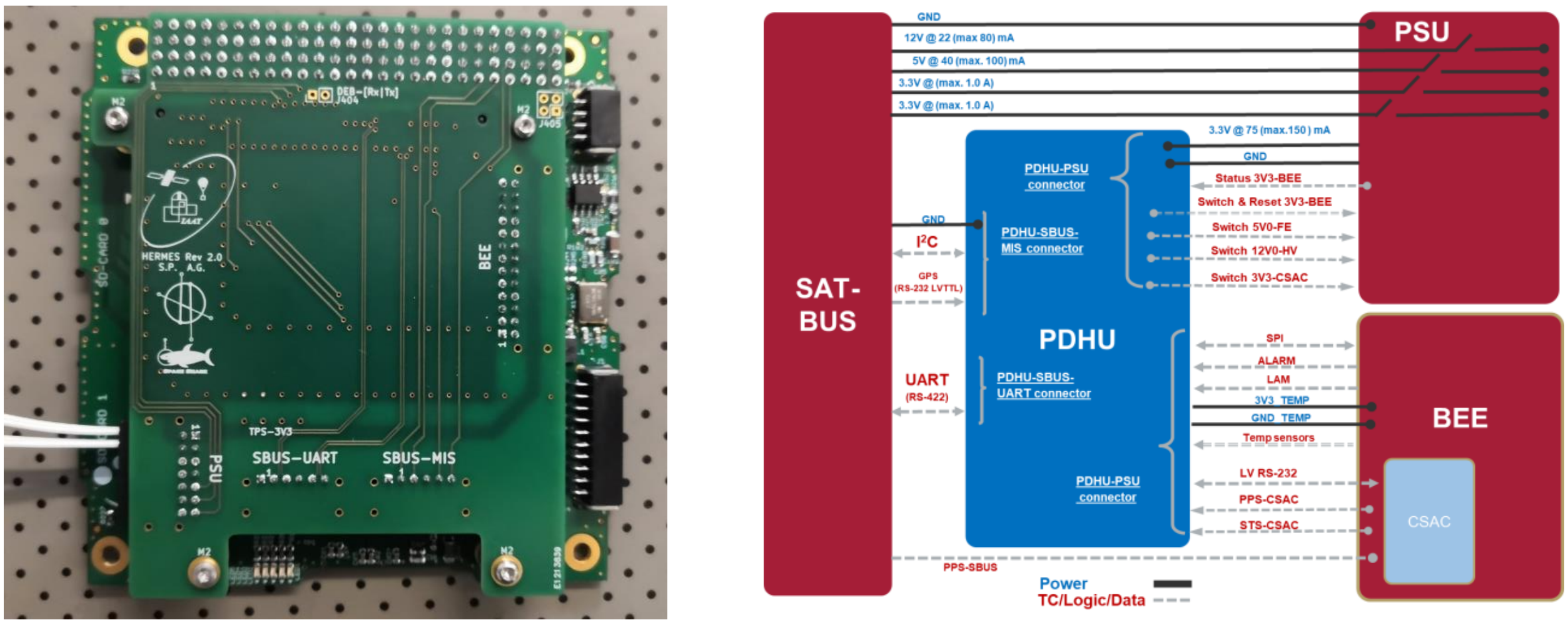

Figure 11 Left: Engineering model of the HERMES-TP/SP Payload Data Handling Unit (PDHU) in the motherboard-daughterboard configuration. Right: Schematic view of the P/L electrical interfaces provided by the PDHU

The electrical interfaces between the PDHU and the spacecraft are obtained by two harnesses connected to the PDHU daughterboard (right panel in Figure 11), one providing the cables for the $I^2C$ protocol (devoted to direct TCs or error/alarm reporting and power cycling the PDHU) and the LVTTL-RS232 UART (used to receive the GPS data), while the other contains the lines for the UART TM/TC communication on a RS-422 bus.

The PDHU provides the payload central processing unit (CPU) and mass memory, and it is in charge of interfacing the PSU for voltage line commanding, the BEE for internal TM/TC transmission and CSAC configuration, and the analog temperature sensors of the payload for temperature monitoring and logging. Moreover, the PDHU manages the payload operative modes, generates and filters the photon list, provides the formatting of the scientific and housekeeping data and perform the burst trigger search. A detailed description of the HERMES-TP/SP PDHU and of its functionalities and performance can be found in [16].

# 4. PAYLOAD PERFORMANCE AND TECHNICAL BUDGETS

In the following, we present the preliminary payload performance as experimentally verified with the payload demonstration model developed during the summer 2020.

## 4.1 Payload demonstration model

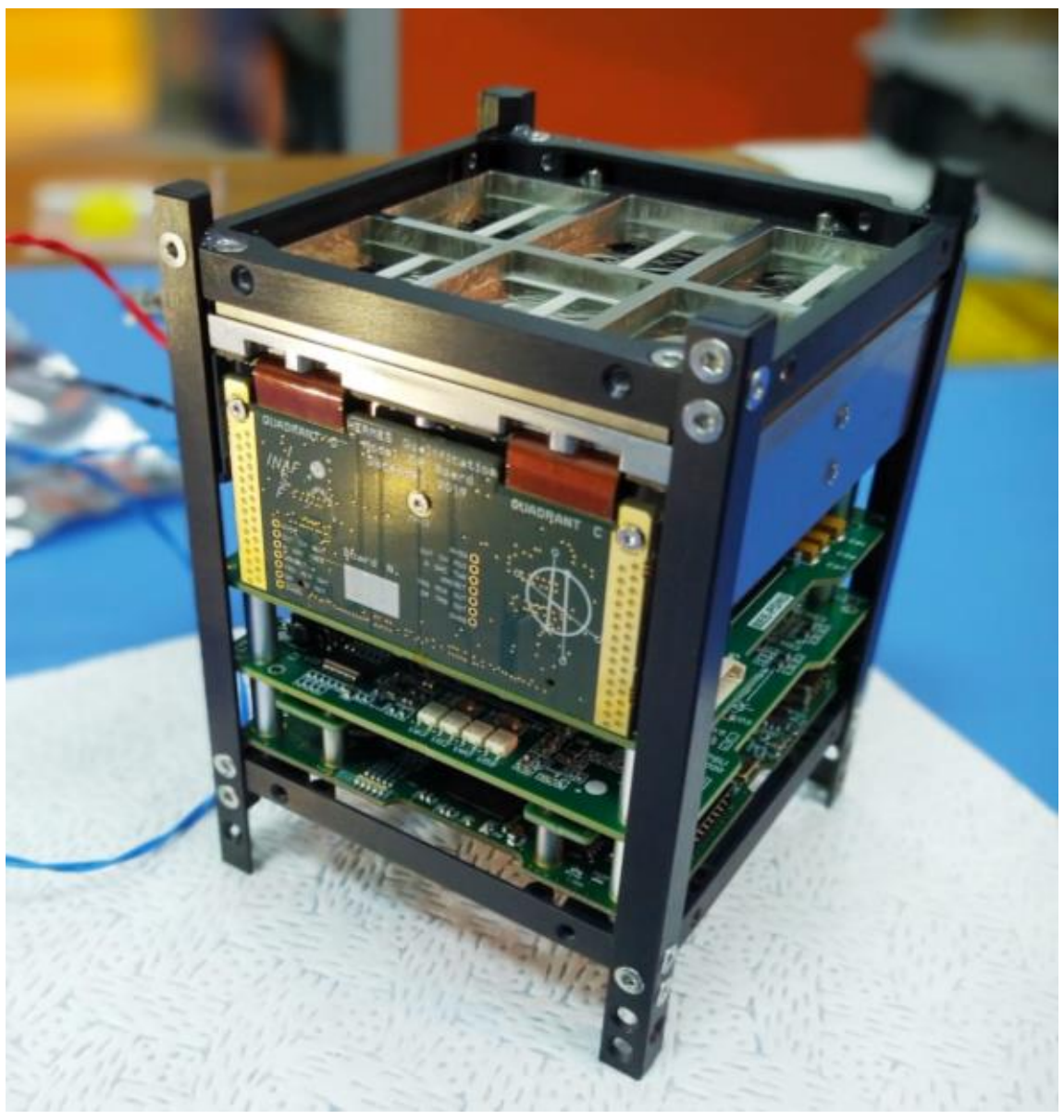

Figure 12 The HERMES-TP/SP PL demonstration model

A complete payload demonstration model (DM) has been built and tested in order to verify that the design of the system fulfills the scientific and system requirements and to validate the AIV&T procedures. The DM has been integrated during the months of August and September 2020 with a representative set of flight-like components, which includes: the payload mechanical assembly completed of tungsten shields, the detector optical filter, the FEE board equipped with 4 LYRA-BE ASICs (one per DA quadrant), 120 LYRA-FE ASICs (30 per quadrant), 4 in-spec SDD arrays correspondent to 40 FEE channels and 8 dummy SDD arrays, 60 GAGG crystals and a fully representative optical coupling and crystal preload pads. The DM (Figure 12) is then completed by the three digital electronics boards (BEE, PSU, PDHU) which are currently undergoing a complete test and characterization campaign.

## 4.2 Detector assembly DM characterization

In order to assess the X-mode and S-mode functionalities and preliminary performance, the integrated DA has been characterized with radioactive sources at the IAPS Rome facility. The DA has been powered by means of unfiltered laboratory power supplies connected, through an adapter board, directly to the LYRA-BE and LYRA-FE ASIC and to the SDD bias network. The LYRA-BE ASICs command and communication tasks have been managed by means of a FPGA-based commercial board (Arty Z7) and a custom developed software and firmware. The output analog signals have been readout using a commercial multi-channel fast ADC (CAEN DT5740) and analyzed using specifically developed algorithms in IDL and Python languages.

A series of energy spectra have been acquired at +26 °C and +5 °C by illuminating the detector simultaneously with X-rays ($^{55}$Fe and $^{109}$Cd) and γ-rays ($^{137}$Cs) sources. Figure 13 reports the distribution of the DM X-mode energy resolution (at 5.9 keV, Mn $K_{\alpha}$ fluorescence line) experimentally measured with a detector leakage current consistent with the end-of-life (EOL) expected current due to the radiation damage. As it can be seen, the system shows an overall noise performance well below the required resolution of 1 keV FWHM EOL reported in Section 2. Moreover, the limited noise allows for a low energy threshold setting lower than 2.5 keV for all the DM channels, as shown in Figure 14.

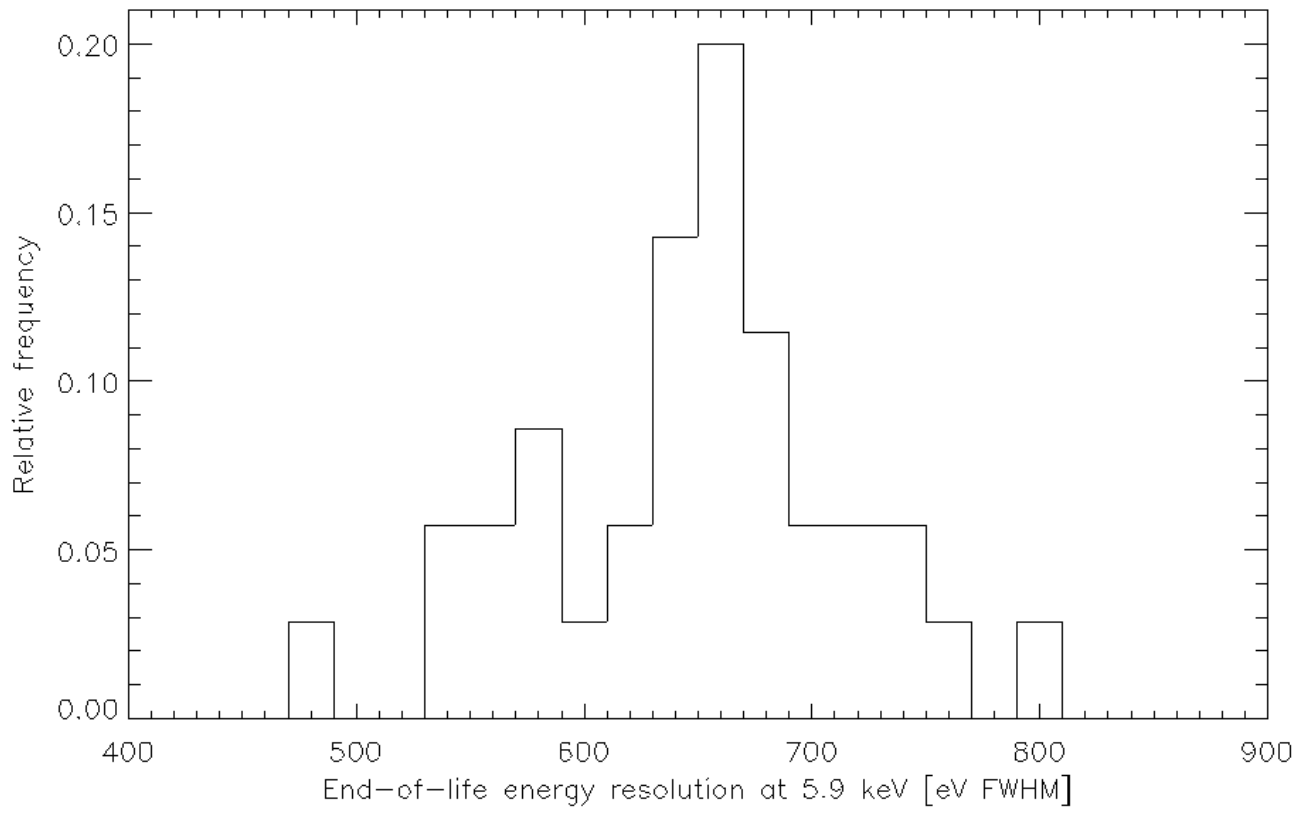

Figure 13 Demonstration model energy resolution at 5.9 keV measured with EOL equivalent detector leakage current

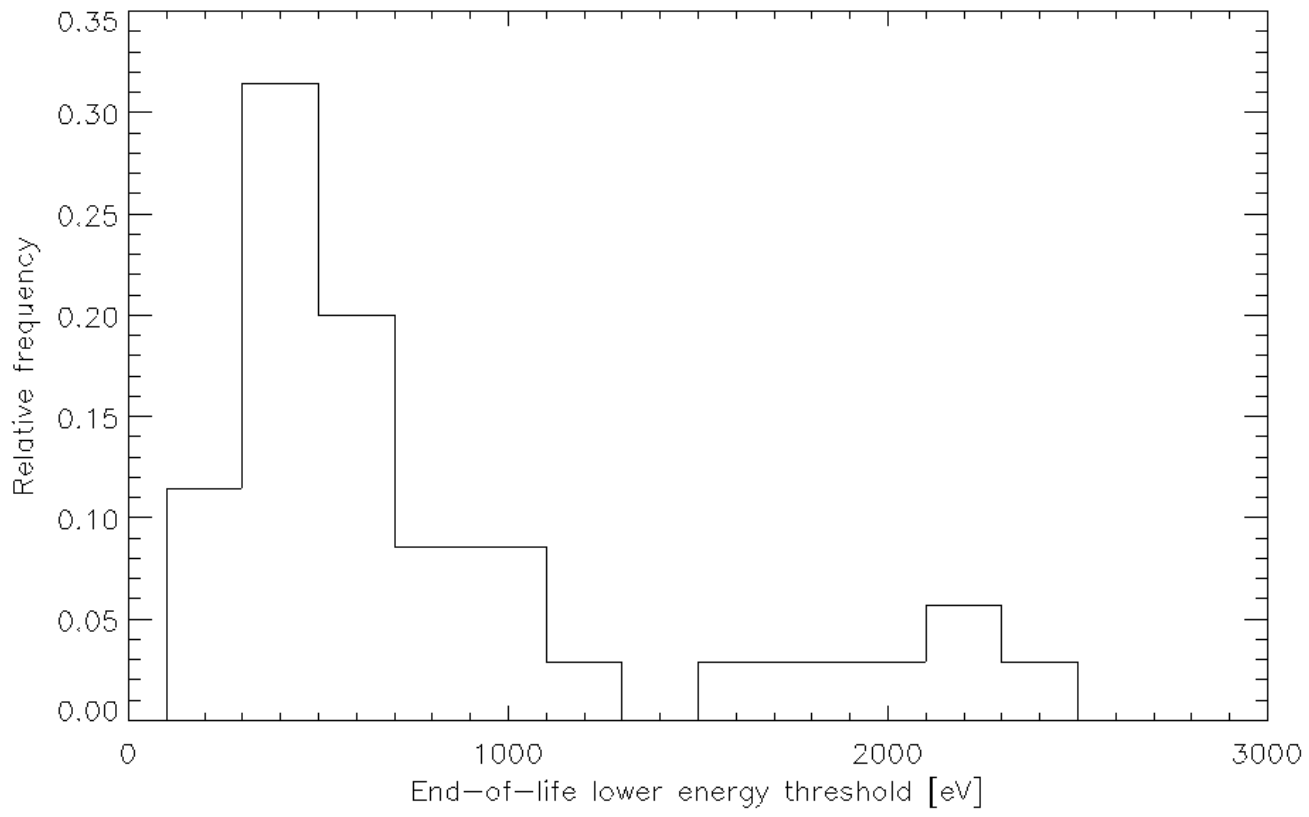

Figure 14 Demonstration model lower energy threshold measured with EOL equivalent detector leakage current

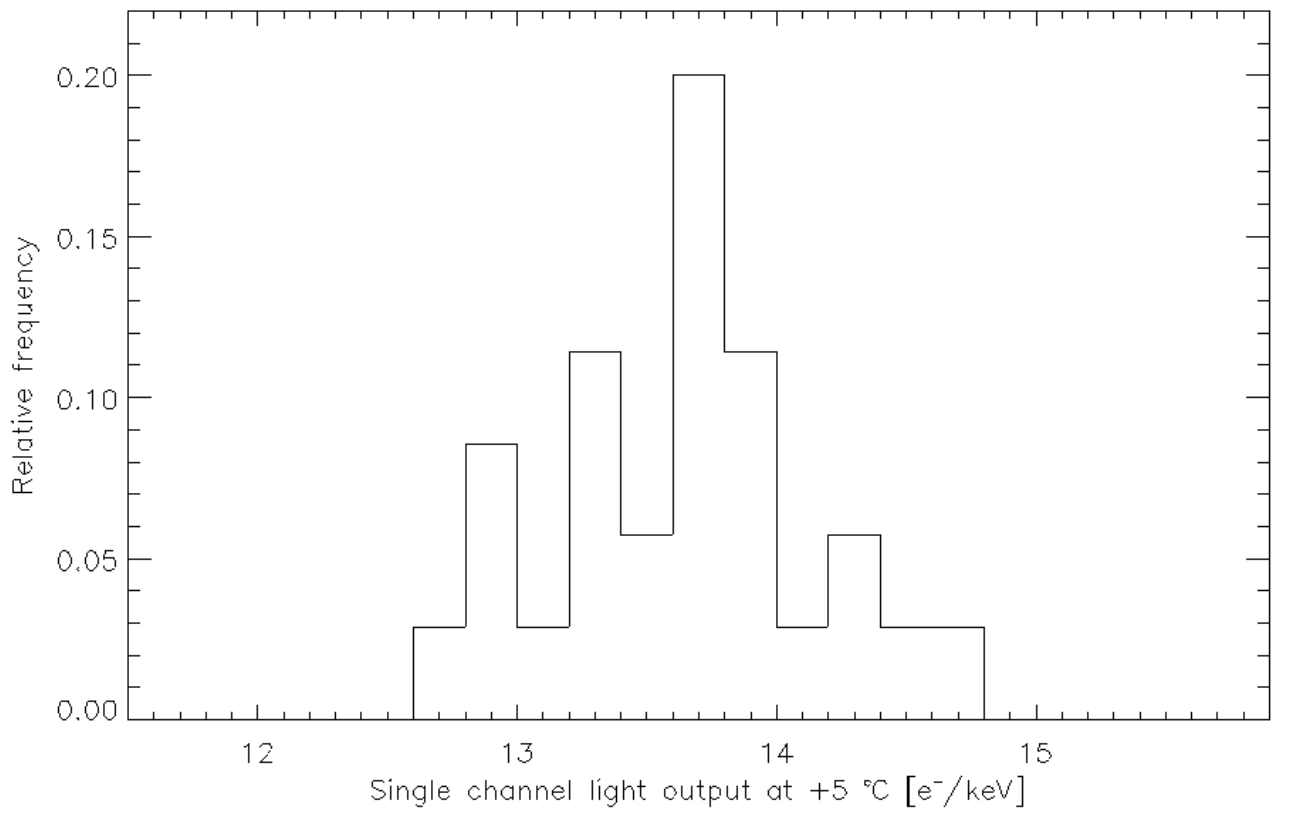

Figure 15 Single channel light output measured at T= +5 °C at 662 keV

Figure 15 shows the distribution of the single channel light output measured with the HERMES-TP/SP DM silicon drift detectors at a temperature of +5° C at 662 keV with a $^{137}Cs$ radioactive source. The average light output is 13.7 $e^-$/keV (which corresponds to 27.4 $e^-$/keV for each Ce:GAGG crystal), with a standard deviation of 0.5 $e^-$/keV. More details about the DM measured performance for both the X-mode and S-mode can be found in [5].

### 4.3 Demonstration model environmental testing

The HERMES-TP/SP P/L mechanical and thermal models have been validated by means of random vibration test and thermal balance test performed at the Politecnico di Milano facilities at the beginning of October 2020.

The DM has been successfully tested with the sine and random vibration levels reported in Table 4 and Table 5 respectively.

Table 4 Sine vibration test levels

| **Axis** | **Frequency [Hz]** | **Qualification level** |
|---|---|---|
| **Out of plane/In plane** | 5–75 | 2.5 g |
| | 75–110 | 1.25 g |
| | 110–125 | 0.25 g |
| **Sweep rate** | | 2 Oct/min 1 sweep up |

Table 5 Random vibration tests levels

| **Frequency [Hz]** | **Level [$G^2$/Hz]** |
|---|---|
| 20 | 0.0224 |
| 50 | 0.0224 |
| 100 | 0.0449 |
| 200 | 0.1122 |
| 500 | 0.1122 |
| 1000 | 0.0449 |
| 2000 | 0.0224 |
| **Overall** | 10.6 $G_{rms}$ |
| **Duration** | 2 minutes |

The thermal balance test (TBT) has been performed to provide data for the verification of the thermal mathematical model as part of the thermal control subsystem qualification, and to provide information about the sensitivity of the thermal control subsystem design with respect to parameter changes. The conditions of the thermal balance set points are reported in Table 6. The number of steady states (6) has been chosen following the ECSS prescriptions.

Table 6 Thermal balance test conditions

| **Test condition** | **Condition** |
|---|---|
| Reference balance | 0 °C |
| Cold balance | –25 °C |
| Hot balance | +30 °C |
| Set point tolerance | ±3 °C |
| Temperature slew rate | <1 °C/min |
| Pressure | $3.5\times10^{-6}$ mBar |
| Thermalization condition | <0.5°C/30min |

Correlation of the temperatures measured during the TBT with the P/L geometrical and thermal mathematical model (G/TMM) showed a temperature mean deviation always lower than 1.6 °C in all the tested states, thus providing validation of the G/TMM.

## 5. CONCLUSIONS

In this paper we presented the design, development and preliminary performance characterization of the HERMES-TP/SP payload, devoted to the study and fast localization of high energy transients with sub-microsecond time resolution and a wide energy band, spanning from few keV up to 2 MeV. A demonstration model has been assembled and tested to validate the payload mechanical and thermal design, as well as to perform the payload requirement verification. The P/L critical design review (CDR) has been successfully passed on October/November 2020. Assembly, integration and test activities (both functional and environmental) of the first payload proto flight module (PFM) is foreseen for Q1/Q2 2021. Production and test of the other five flight modules (FMs) will follow, ending in Q4 2021.

Table 7 summarizes the technical and performance budgets of the HERMES-TP/SP P/L as measured during the characterization of the payload demonstration model, while detailed studies of the HERMES-TP/SP space environment and of the instrument in-flight performance can be found in [17], [18], [19] and [20].

Table 7 Technical and performance budget of the HERMES-TP/SP payload

| **Parameter** | **Value** |
|---|---|
| Payload peak effective area (X-mode & S-mode) | 52 $cm^2$ |
| Field of View | 3.2 sr FWHM |
| Lower energy threshold | ≤ 3 keV |
| Energy resolution X-mode (@ 6 keV) | ≤ 800 eV FWHM |
| Energy resolution S-mode (@ 60 keV) | ≤ 5 keV FWHM |
| Time resolution (68% c.l.) | 320 ns |
| Time accuracy (68% c.l.) | 181 ns |
| Payload mass | 1588 g |
| Payload power in Observation | 4791 mW |
| Payload Telemetry (scientific + engineering) | 719 Mbits/day |

## ACKNOWLEDGMENTS

This work has been carried out in the framework of the HERMES-TP and HERMES-SP collaborations. We acknowledge support from the European Union Horizon 2020 Research and Innovation Framework Programme under grant agreement HERMES-Scientific Pathfinder n. 821896 and from ASI-INAF Accordo Attuativo HERMES Technologic Pathfinder n. 2018-10-H.1-2020.